\documentclass[twocolumn,showpacs,preprintnumbers,superscriptaddress,amsmath,amssymb,pre]{revtex4}

\usepackage{graphicx}
\usepackage{dcolumn}
\usepackage{bm}
\usepackage{color}

\begin{document}

\preprint{PRE/2D-(MF)-DFA}

\title{Detrended fluctuation analysis for fractals and multifractals in higher dimensions}

\author{Gao-Feng Gu}
 \affiliation{School of Business, East China University of Science and Technology, Shanghai 200237, China} %
 \affiliation{Research Center of Systems Engineering, East China University of Science and Technology, Shanghai 200237, China} %

\author{Wei-Xing Zhou}
 \email{wxzhou@moho.ess.ucla.edu}
 \affiliation{School of Business, East China University of Science and Technology, Shanghai 200237, China} %
 \affiliation{Research Center of Systems Engineering, East China University of Science and Technology, Shanghai 200237, China} %
 \affiliation{School of Science, East China University of Science and Technology, Shanghai 200237, China} %

\date{\today}

\begin{abstract}
One-dimensional detrended fluctuation analysis (1D DFA) and
multifractal detrended fluctuation analysis (1D MF-DFA) are widely
used in the scaling analysis of fractal and multifractal time series
because of being accurate and easy to implement. In this paper we
generalize the one-dimensional DFA and MF-DFA to higher-dimensional
versions. The generalization works well when tested with synthetic
surfaces including fractional Brownian surfaces and multifractal
surfaces. The two-dimensional MF-DFA is also adopted to analyze two
images from nature and experiment and nice scaling laws are
unraveled.
\end{abstract}


\maketitle


\section{Introduction}
\label{s1:Introduction}

Fractals and multifractals are ubiquitous in natural and social
sciences \cite{Mandelbrot-1983}.  The most usual records of
observable quantities are in the form of time series and their
fractal and multifractal properties have been extensively
investigated. There are many methods proposed for this purpose
\cite{Taqqu-Teverovsky-Willinger-1995-Fractals,Montanari-Taqqu-Teverovsky-1999-MCM},
such as spectral analysis, rescaled range analysis
\cite{Hurst-1951-TASCE,Mandelbrot-Ness-1968-SIAMR,Mandelbrot-Wallis-1969a-WRR,Mandelbrot-Wallis-1969b-WRR,Mandelbrot-Wallis-1969c-WRR,Mandelbrot-Wallis-1969d-WRR},
fluctuation analysis
\cite{Peng-Buldyrev-Goldberger-Havlin-Sciortino-Simons-Stanley-1992-Nature},
detrended fluctuation analysis
\cite{Peng-Buldyrev-Havlin-Simons-Stanley-Goldberger-1994-PRE,Hu-Ivanov-Chen-Carpena-Stanley-2001-PRE,Kantelhardt-Zschiegner-Bunde-Havlin-Bunde-Stanley-2002-PA},
wavelet transform module maxima (WTMM)
\cite{Holschneider-1988-JSP,Muzy-Bacry-Arneodo-1991-PRL,Muzy-Bacry-Arneodo-1993-JSP,Muzy-Bacry-Arneodo-1993-PRE,Muzy-Bacry-Arneodo-1994-IJBC},
detrended~moving~average
\cite{Alessio-Carbone-Castelli-Frappietro-2002-EPJB,Carbone-Castelli-Stanley-2004-PA,Carbone-Castelli-Stanley-2004-PRE,Alvarez-Ramirez-Rodriguez-Echeverria-2005-PA,Xu-Ivanov-Hu-Chen-Carbone-Stanley-2005-PRE},
to list a few. It is now the common consensus that DFA and WTMM have
the highest precision in the scaling analysis
\cite{Taqqu-Teverovsky-Willinger-1995-Fractals,Montanari-Taqqu-Teverovsky-1999-MCM,Audit-Bacry-Muzy-Arneodo-2002-IEEEtit}.

The idea of DFA was invented originally to investigate the
long-range dependence in coding and noncoding DNA nucleotides
sequence
\cite{Peng-Buldyrev-Havlin-Simons-Stanley-Goldberger-1994-PRE}. Then
it was generalized to study the multifractal nature hidden in time
series, termed as multifractal DFA (MF-DFA)
\cite{Kantelhardt-Zschiegner-Bunde-Havlin-Bunde-Stanley-2002-PA}.
Due to the simplicity in implementation, the DFA is now becoming the
most important method in the field.

Although the WTMM method seems a little bit complicated, it is no
doubt a very powerful method, especially for high-dimensional
objects, such as images, scalar and vector fields of
three-dimensional turbulence
\cite{Arneodo-Decoster-Roux-2000-EPJB,Decoster-Roux-Arneodo-2000-EPJB,Roux-Arneodo-Decoster-2000-EPJB,Kestener-Arneodo-2003-PRL,Kestener-Arneodo-2004-PRL}.
In contrast, the original DFA method is not designed for such
purpose. In a recent paper, a first effort is taken to apply DFA to
study the roughness features of texture images
\cite{AlvarezRamirez-Rodriguez-Cervantes-Carlos-2006-PA}.
Specifically, the DFA is applied to extract Hurst indices of the
one-dimensional sequences at different image orientations and their
average scaling exponent is estimated. Unfortunately, this is
nevertheless a one-dimensional DFA method.

In this work, we generalize the DFA (and MF-DFA as well) method from
one-dimensional to high-dimensional. The generalized methods are
tested by synthetic surfaces (fractional Brownian surfaces and
multifractal surfaces) with known fractal and multifractal
properties. The numerical results are in excellent agreement with
the theoretical properties. We then apply these methods to practical
examples. We argue that there are tremendous potential applications
of the generalized DFA to many objects, such as the roughness of
fracture surfaces, landscapes, clouds, three-dimensional temperature
fields and concentration fields, turbulence velocity fields.

The paper is organized as follows. In Sec.~\ref{s1:Method}, we
represent the algorithm of the two-dimensional detrended fluctuation
analysis and the two-dimensional multifractal detrended fluctuation
analysis. Section~\ref{s1:Simulation} shows the results of the
numerical simulations, which are compared with theoretical
properties. Applications to practical examples are illustrated in
Sec.~\ref{s1:Example}. We discuss and conclude in
Sec.~\ref{s1:Conclusion}.

\section{Method} \label{s1:Method}

\subsection{Two-dimensional DFA}
\label{s2:Method:DFA}

Being a direct generalization, the higher-dimensional DFA and MF-DFA
have quite similar procedures as the one-dimensional DFA. We shall
focus on two-dimensional space and the generalization to
higher-dimensional is straightforward. The two-dimensional DFA
consists of the following steps.

Step 1: Consider a self-similar (or self-affine) surface, which is
denoted by a two-dimensional array $X(i,j)$, where $i=1,2,\cdots,M$,
and $j = 1, 2,\cdots, N$. The surface is partitioned into $M_s
\times N_s$ disjoint square segments of the same size $s\times s$,
where $M_s = [M/s]$ and $N_s = [N/s]$. Each segment can be denoted
by $X_{v,w}$ such that $X_{v,w}(i,j)=X(l_1+i,l_2+j)$ for
$1\leqslant{i,j}\leqslant{s}$, where $l_1=(v-1)s$ and $l_2=(w-1)s$.

Step 2: For each segment $X_{v,w}$ identified by $v$ and $w$, the
cumulative sum $u_{v,w}(i,j)$ is calculated as follows:
\begin{equation}
u_{v,w}(i,j) = \sum_{k_1=1}^{i}\sum_{k_2=1}^{j}{X_{v,w}(k_1,k_2)}~,
 \label{Eq:DFA_u}
\end{equation}
where $1\leqslant{i,j}\leqslant{s}$. Note that $u_{v,w}$ itself is a
surface.

Step 3: The trend of the constructed surface $u_{v,w}$ can be
determined by fitting it with a prechosen bivariate polynomial
function $\widetilde{u}$. The simplest function could be a plane. In
this work, we shall adopt the following five detrending functions to
test the validation of the methods:
\begin{eqnarray}
\widetilde{u}_{v,w}(i,j)&=&ai+bj+c~,\label{Eq:eq30}\\
\widetilde{u}_{v,w}(i,j)&=&ai^2+bj^2+c~,\label{Eq:eq31}\\
\widetilde{u}_{v,w}(i,j)&=&aij+ bi+cj+d~,\label{Eq:eq4}\\
\widetilde{u}_{v,w}(i,j)&=&ai^2+bj^2+ci+dj+e~,\label{Eq:eq5}\\
\widetilde{u}_{v,w}(i,j)&=&ai^2+bj^2+cij+di+ej+f~,\label{Eq:eq6}
\end{eqnarray}
where $1\leqslant{i,j}\leqslant{s}$ and $a$, $b$, $c$, $d$, $e$, and
$f$ are free parameters to be determined. These parameters can be
estimated easily through simple matrix operations, derived from the
least squares method. We can then obtain the residual matrix:
\begin{equation}
\epsilon_{v,w}(i,j)=u_{v,w}(i,j)-\widetilde{u}_{v,w}(i,j)~.
\end{equation}
The detrended fluctuation function $F(v,w,s)$ of the segment
$X_{v,w}$ is defined via the sample variance of the residual matrix
$\epsilon_{v,w}(i,j)$ as follows:
\begin{equation}
F^2(v,w,s) = \frac{1}{s^2}\sum_{i = 1}^{s}\sum_{j =
1}^{s}\epsilon_{v,w}(i,j)^2~.
 \label{Eq:DFA_F}
\end{equation}
Note that the mean of the residual is zero due to the detrending
procedure.

Step 4: The overall detrended fluctuation is calculated by averaging
over all the segments, that is,
\begin{equation}
F^2(s) =\frac{1}{M_sN_s}\sum_{v = 1}^{M_s}\sum_{w =
1}^{N_s}{F^2(v,w,s)}~.
 \label{Eq:DFA_Fs}
\end{equation}

Step 5: Varying the value of $s$ in the range from $s_{\min} \approx
6$ to $s_{\max} \approx \min(M,N)/4$, we can determine the scaling
relation between the detrended fluctuation function $F(s)$ and the
size scale $s$, which reads
\begin{equation}
F(s) \sim s^{H}~,
 \label{Eq:DFA_2D_H}
\end{equation}
where $H$ is the Hurst index of the surface
\cite{Taqqu-Teverovsky-Willinger-1995-Fractals,Kantelhardt-Bunde-Rego-Havlin-Bunde-2001-PA,Talkner-Weber-2000-PRE,Heneghan-McDarby-2000-PRE},
which can be related to the fractal dimension by $D=3-H$
\cite{Mandelbrot-1983,Voss-1989-PD}.

Since $N$ and $M$ need not be a multiple of the segment size $s$,
two orthogonal trips at the end of the profile may remain. In order
to take these ending parts of the surface into consideration, the
same partitioning procedure can be repeated starting from the other
three corners \cite{Kantelhardt-Bunde-Rego-Havlin-Bunde-2001-PA}.

\subsection{Two-dimensional MF-DFA}
\label{s2:Method:MFDFA}

Analogous to the generalization of one-dimensional DFA to
one-dimensional MF-DFA, the two-dimensional MF-DFA can be ascribed
similarly, such that the two-dimensional DFA serves as a special
case of the two-dimensional MF-DFA. The two-dimensional MF-DFA
follows the same first three steps as in the two-dimensional DFA and
has two revised steps.

Divide a self-similar (or self-affine) surface $X(i,j)$ into $M_s
\times N_s$ ($M_s = [M/s]$ and $N_s = [N/s]$) disjoint phalanx
segments. In each segment $X_{v,w}(i,j)$ compute the cumulative sum
$u(i,j,s)$ using Eq.~(\ref{Eq:DFA_u}). With one of the five
regression equations, we can obtain ${\widetilde{u}(i,j,s)}$ to
represent the trend in each segment, then we obtain the fluctuate
function $F(v,w,s)$ by Eq.~(\ref{Eq:DFA_F}).

Step 4: The overall detrended fluctuation is calculated by averaging
over all the segments, that is,
\begin{equation}
F_q(s) = \left\{\frac{1}{M_sN_s}\sum_{v = 1}^{M_s}\sum_{w =
1}^{N_s}{[F(v,w,s)]^q}\right\}^{1/q}~,
 \label{Eq:DFA_M_Fqs}
\end{equation}
where $q$ can take any real value except for $q = 0$. When $q = 0$,
we have
\begin{equation}
F_0(s) = \exp\left\{\frac{1}{M_sN_s}\sum_{v = 1}^{M_s}\sum_{w =
1}^{N_s}{\ln[F(v,w,s)]}\right\}~,
 \label{Eq:DFA_M_Fq0}
\end{equation}
according to L'H\^{o}spital's rule.

Step 5: Varying the value of $s$ in the range from $s_{\min} \approx
6$ to $s_{\max} \approx \min(M,N)/4$, we can determine the scaling
relation between the detrended fluctuation function $F_q(s)$ and the
size scale $s$, which reads
\begin{equation}
F_q(s) \sim s^{h(q)}~.
 \label{Eq:DFA_M_h}
\end{equation}

For each $q$, we can get the corresponding traditional ${\tau(q)}$
function through
\begin{equation}
\tau(q) = qh(q) - D_f~, \label{Eq:MFDFA:tau}
\end{equation}
where $D_f$ is the fractal dimension of the geometric support of the
multifractal measure
\cite{Kantelhardt-Zschiegner-Bunde-Havlin-Bunde-Stanley-2002-PA}. It
is thus easy to obtain the generalized dimensions $D_q$
\cite{Grassberger-1983-PLA,Hentschel-Procaccia-1983-PD,Grassberger-Procaccia-1983-PD}
and the singularity strength function $\alpha(q)$ and the
multifractal spectrum $f(\alpha)$ via Legendre transform
\cite{Halsey-Jensen-Kadanoff-Procaccia-Shraiman-1986-PRA}. In this
work, the numerical and real multifractals have $D_f=2$. For
fractional Brownian surfaces with a Hurst index $H$, we have $h(q)
\equiv H$.

\subsection{A note on the generalization}
\label{s2:Method:Note}

To the best of our knowledge, the first few steps of the
one-dimensional DFA and MF-DFA in literature are organized in the
following order: Construct the cumulative sum of the time series and
then partition it into segments of the same scale without
overlapping. In this way, a direct generalization to higher
dimensional space should be the following:

Step I: Construct the cumulative sum
\begin{equation}
 u(i,j) = \sum_{k_1 = 1}^{i}\sum_{k_2 = 1}^{j}{X(k_1,k_2)},
 \label{Eq:III:c_u}
\end{equation}

Step II: Partition $u(i,j)$ into $N_s\times{M_s}$ disjoint square
segments. The ensuing steps are the same as those described in
Sec.~\ref{s2:Method:DFA} and Sec.~\ref{s2:Method:MFDFA}.

It is easy to show that, for the one-dimensional DFA and MF-DFA, the
residual matrix in a given segment is the same no matter which step
is processed first, either the cumulative summation or the
partitioning. This means that we have two manners of generalizing to
higher-dimensional space, that is, Steps 1-2 in
Sec.~\ref{s2:Method:DFA} and Steps I-II aforementioned. Our
numerical simulations show that both these two kinds of
generalization gives the correct Hurst index for fractional Brownian
surfaces when adopting two-dimensional DFA. However, the
two-dimensional MF-DFA with Steps I-II gives wrong $\tau(q)$
function for two-dimensional multifractals with analytic solutions
where the power-law scaling is absent, while the generalization with
Steps 1-2 does a nice job.

The difference of the two generalization methods becomes clear when
we compare $u_{v,w}(i,j)$ in Eq.~(\ref{Eq:DFA_u}) with $u(i,j)$ in
Eq.~(\ref{Eq:III:c_u}). We see that $u_{v,w}(l_1+i,l_2+j)$ is
localized to the segment $X_{v,w}$, while $u(l_1+i,l_2+j)$ contains
extra information outside the segment when $i<l_1$ and $j<l_2$,
which is not constant for different $i$ and $j$ and thus can not be
removed by the detrending procedure. In the following sections, we
shall therefore concentrate on the correct generalization expressed
in Sec.~\ref{s2:Method:DFA} and Sec.~\ref{s2:Method:MFDFA}.

\section{Numerical simulations}
\label{s1:Simulation}

\subsection{Synthetic fractional Brownian surfaces}

We test the two-dimensional DFA with synthetic fractional Brownian
surfaces. There are many different methods to create fractal
surfaces, based on Fourier transform filtering
\cite{Peitgen-Saupe-1988,Voss-1989-PD}, midpoint displacement and
its variants
\cite{Mandelbrot-1983,Fournier-Fussell-Carpenter-1982-CACM,Koh-Hearn-1992-CGF},
circulant embedding of covariance matrix
\cite{Dietrich-Newsam-1993-WRR,Dietrich-Newsam-1997-SIAMjsc,Wood-Chan-1994-JCGS,Chan-Wood-1997-JRSSC},
periodic embedding and fast Fourier transform
\cite{Stein-2002-JCGS}, top-down hierarchical model
\cite{Penttinen-Virtamo-2004-MCAP}, and so on. In this paper, we use
the free MATLAB software FracLab 2.03 developed by INRIA to
synthesize fractional Brownian surfaces with Hurst index $H$.

\begin{figure}[htb]
\begin{center}
\includegraphics[width=6.5cm]{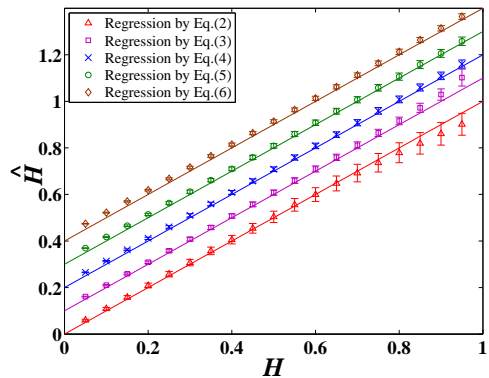}
\caption{Comparison of the estimated Hurst index $\hat{H}$ using
Eqs.~(\ref{Eq:eq30}-\ref{Eq:eq6}) with the true value $H$. The error
bars show the standard deviation of the 500 estimated $\hat{H}$
values. The results corresponding to
Eqs.~(\ref{Eq:eq31}-\ref{Eq:eq6}) are translated vertically for
clarity.} \label{Fig:fBM:DFA:H}
\end{center}
\end{figure}

In our test, we have investigated fractional Brownian surfaces with
different Hurst indices $H$ ranging from $0.05$ to $0.95$ with an
increment of $0.05$. The size of the simulated surfaces is $500
\times 500$. For each $H$, we generated $500$ surfaces. Each surface
is analyzed by the two-dimensional DFA with the five bivariate
functions in Eqs.~(\ref{Eq:eq30}-\ref{Eq:eq6}). The results are
shown in Fig.~\ref{Fig:fBM:DFA:H}. We can see that the estimated
Hurst indexes $\hat{H}$ are very close to the preset values in
general. The deviation of Hurst index $H$ becomes larger for large
values of $H$.

\begin{figure}[htb]
\begin{center}
\includegraphics[width=6.5cm]{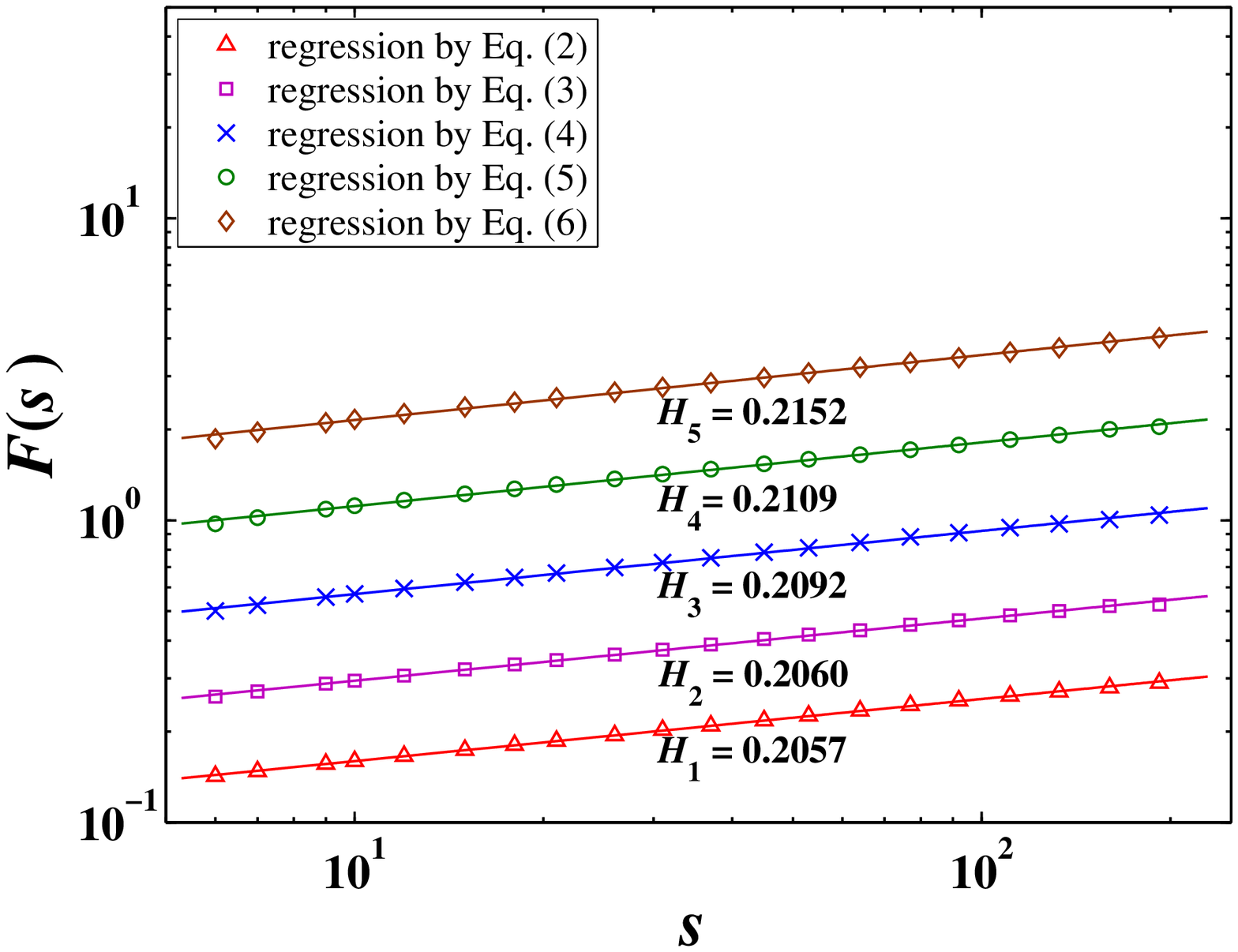}
\includegraphics[width=6.5cm]{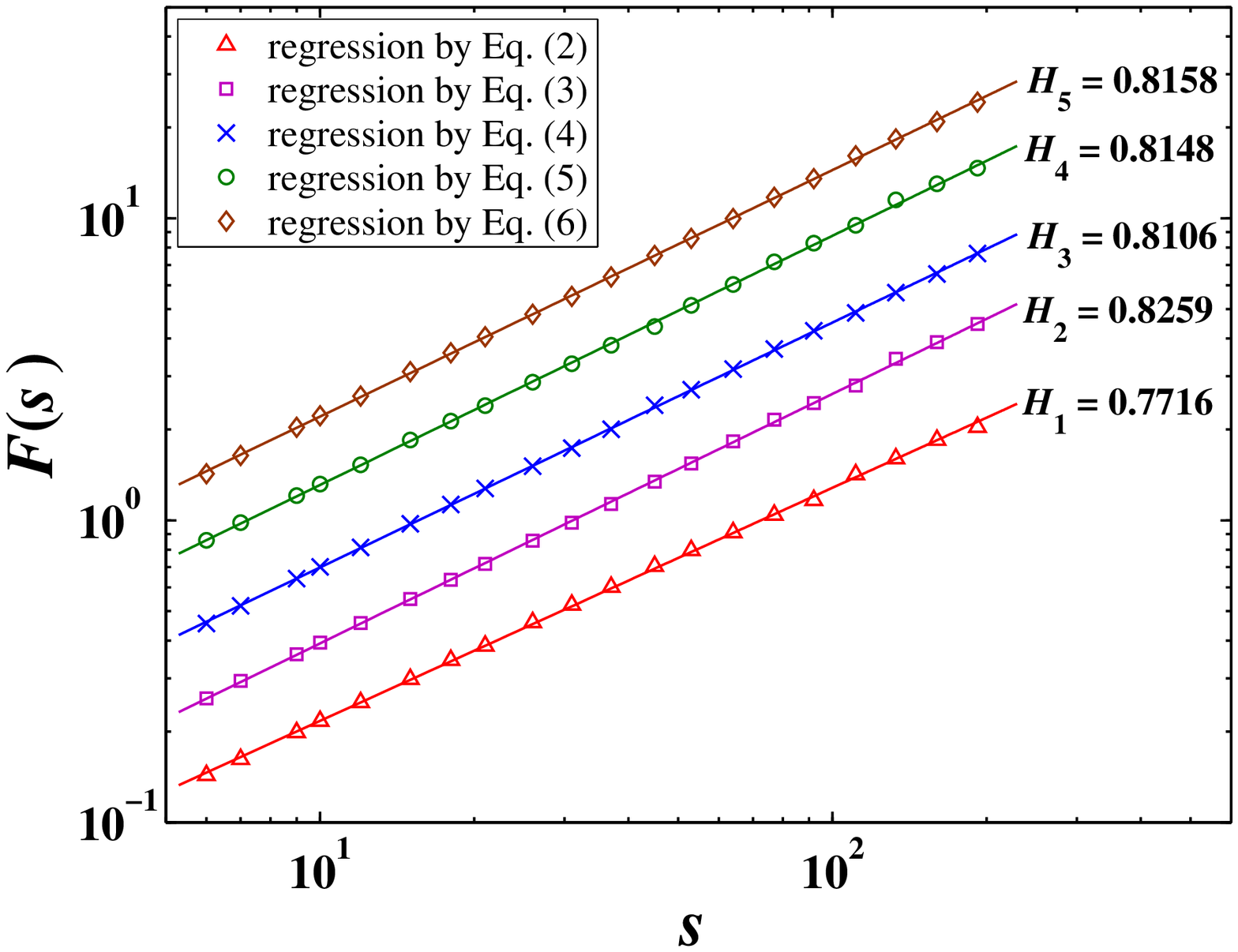}
\caption{Log-log plots of the detrended fluctuation function $F(s)$
with respect to the scale $s$ for $H = 0.2$ (top panel) and $H =
0.8$ (bottom panel) using Eqs.~(\ref{Eq:eq30}-\ref{Eq:eq6}). The
lines are the least squares fits to the data. The results
corresponding to Eqs.~(\ref{Eq:eq31}-\ref{Eq:eq6}) are translated
vertically for clarity.} \label{Fig:fBM:DFA:Fs}
\end{center}
\end{figure}

In Fig.~\ref{Fig:fBM:DFA:Fs}, we show the log-log plot of the
detrended fluctuation $F(s)$ as a function of $s$ for two synthetic
fractional Brownian surfaces with $H = 0.2$ and $H = 0.8$,
respectively. There is no doubt that the power-law scaling between
$F(s)$ and $s$ is very evident and sound. Hence, the two-dimensional
DFA is able to well capture the self-similar nature of the
fractional Brownian surfaces and results in precise estimation of
the Hurst index.

\begin{figure}[htb]
\begin{center}
\includegraphics[width=6.5cm]{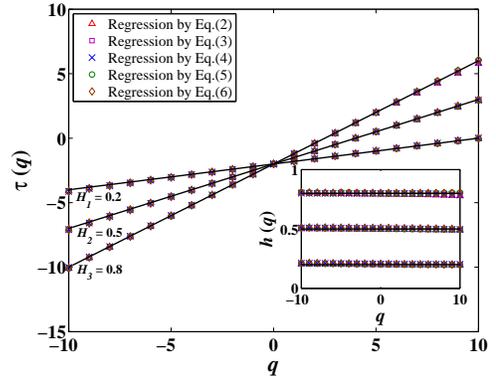}
\caption{Plots of $\tau(q)$ extracted by using the five detrending
functions (\ref{Eq:eq30}-\ref{Eq:eq6}) as a function of $q$. The
three straight lines are $\tau(q)=qH-2$ for $H_1=0.2$, $H_2=0.5$,
and $H_3=0.8$, respectively. The inset shows the corresponding
$h(q)$ functions.} \label{Fig:fBM:MFDFA}
\end{center}
\end{figure}

We also adopted fractional Brownian surfaces to test the
two-dimensional multifractal detrended fluctuation analysis.
Specifically, we have simulated three fractional Brownian surfaces
with Hurst indexes $H_1=0.2$, $H_2=0.5$, and $H_3=0.8$,
respectively. The five regression equations
(\ref{Eq:eq30}-\ref{Eq:eq6}) are used in the detrending. We
calculated $h(q)$ for $q$ ranging from $-10$ to $10$ according to
Eq.~(\ref{Eq:DFA_M_h}). All the $F_q(s)$ functions exhibit excellent
power-law scaling with respect to the scale $s$. The function
$\tau(q)$ can be determined according to Eq.~(\ref{Eq:MFDFA:tau}).
The resultant $\tau(q)$ functions are plotted in
Fig.~\ref{Fig:fBM:MFDFA} with the inset showing the $h(q)$
functions. We can find from the figure that, for each surface, the
five functions of $\tau(q)$ (and $h(q)$ as well) corresponding to
the five detrending functions collapse on a single curve. Moreover,
it is evident that $h(q)=H$ and $\tau(q)=qH-2$. The three analytic
straight lines intersect at the same point $(q=0,\tau(q)=-2)$. These
results are expected according to theoretical analysis.

We stress that, when fractional Brownian surfaces are under
investigation, both the two-dimensional DFA and MF-DFA can produce
the same correct results even when Steps I-II are adopted.

\subsection{Synthetic two-dimensional multifractals}

Now we turn to test the MF-DFA method with synthetic two-dimensional
multifractal measures. There exits several methods for the synthesis
of two-dimensional multifractal measures or multifractal rough
surfaces \cite{Decoster-Roux-Arneodo-2000-EPJB}. The most classic
method follows a multiplicative cascading process, which can be
either deterministic or stochastic
\cite{Mandelbrot-1974-JFM,Meneveau-Sreenivasan-1987-PRL,Novikov-1990-PFA,Meneveau-Sreenivasan-1991-JFM}.
The simplest one is the $p$-model proposed to mimick the kinetic
energy dissipation field in fully developed turbulence
\cite{Meneveau-Sreenivasan-1987-PRL}. Starting from a square, one
partitions it into four sub-squares of the same size and chooses
randomly two of them to assign the measure of $p/2$ and the
remaining two of $(1-p)/2$. This partitioning and redistribution
process repeats and we obtain a singular measure $\mu$. A
straightforward derivation following the partition function method
\cite{Halsey-Jensen-Kadanoff-Procaccia-Shraiman-1986-PRA} results in
the analytic expression:
\begin{equation}\label{Eq:MFS1:tau}
    \tau(q) = q-1-\log_2\left[p^q+(1-p)^q\right]~.
\end{equation}

A relevant method is the fractionally integrated singular cascade
(FISC) method, which was proposed to model multifractal geophysical
fields \cite{Schertzer-Lovejoy-1987-JGR} and turbulent fields
\cite{Schertzer-Lovejoy-Schmitt-Chigirinskaya-Marsan-1997-Fractals}.
The FISC method consists of a straightforward filtering in Fourier
space via fractional integration of a singular multifractal measure
generated with some multiplicative cascade process so that the
multifractal measure is transformed into a smoother multifractal
surface:
\begin{equation}\label{Eq:MFS2:conv}
    f(x)=\mu(x)\otimes|x|^{-(1-H)}~,
\end{equation}
where $\otimes$ is the convolution operator and $H\in(0,1)$ is the
order of the fractional integration
\cite{Decoster-Roux-Arneodo-2000-EPJB}, whose $\tau(q)$ function is
\cite{Arrault-Arneodo-Davis-Marshak-1997-PRL,Decoster-Roux-Arneodo-2000-EPJB}:
\begin{equation}\label{Eq:MFS2:tau}
    \tau(q) = q(1+H)-1-\log_2\left[p^q+(1-p)^q\right]~.
\end{equation}
The third one is called the random $\cal{W}$ cascade method which
generates multifractal rough surfaces from random cascade process on
separable wavelet orthogonal basis
\cite{Decoster-Roux-Arneodo-2000-EPJB}.

\begin{figure}[htb]
\begin{center}
\includegraphics[width=6.5cm]{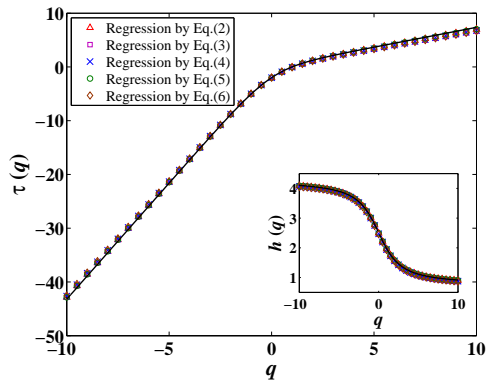}
\caption{Plots of $\tau(q)$ extracted by using the five detrending
functions (\ref{Eq:eq30}-\ref{Eq:eq6}) as a function of $q$. The
continuous line is the theoretical formula (\ref{Eq:MFS3:tau}). The
inset shows the corresponding $h(q)$ functions.}
\label{Fig:MfS:MFDFA}
\end{center}
\end{figure}

In our test, we adopted the first method for the synthesis of
two-dimensional multifractal measure. Starting from a square, one
partitions it into four sub-squares of the same size and assigns
four given proportions of measure $p_1=0.05$, $p_2=0.15$,
$p_3=0.20$, and $p_4=0.60$ to them. Then each sub-square is divided
into four smaller squares and the measure is redistributed in the
same way. This procedure is repeated 10 times and we generate
multifractal ``surfaces'' of size $1024 \times 1024$. The resultant
$\tau(q)$ functions estimated from the two-dimensional MF-DFA method
are plotted in Fig.~\ref{Fig:MfS:MFDFA}, where the inset shows the
$h(q)$ functions. We can find that the five functions of $\tau(q)$
(and $h(q)$ as well) corresponding to the five detrending functions
collapse on a single curve, which is in excellent agreement with the
theoretical formula:
\begin{equation}\label{Eq:MFS3:tau}
    \tau(q) = -\log_2\left(p_1^q+p_2^q+p_3^q+p_4^q\right)~.
\end{equation}

We stress that, when we use Steps I-II instead of Steps 1-2, the
resultant $\tau(q)$ estimated by the MF-DFA method deviates
remarkably from the theoretical formula. Indeed, the power-law
scaling for most $q$ values is absent and thus the alternative
algorithm with Steps I-II and the resulting $\tau(q)$ is completely
wrong. In addition, we see that different detrending functions give
almost the same results. The linear function (\ref{Eq:eq30}) is
preferred in practice, since it requires the least computational
time among the five.

\section{Examples of image analysis}
\label{s1:Example}

\subsection{The data}

\begin{figure}[htb]
\begin{center}
\includegraphics[width=4cm]{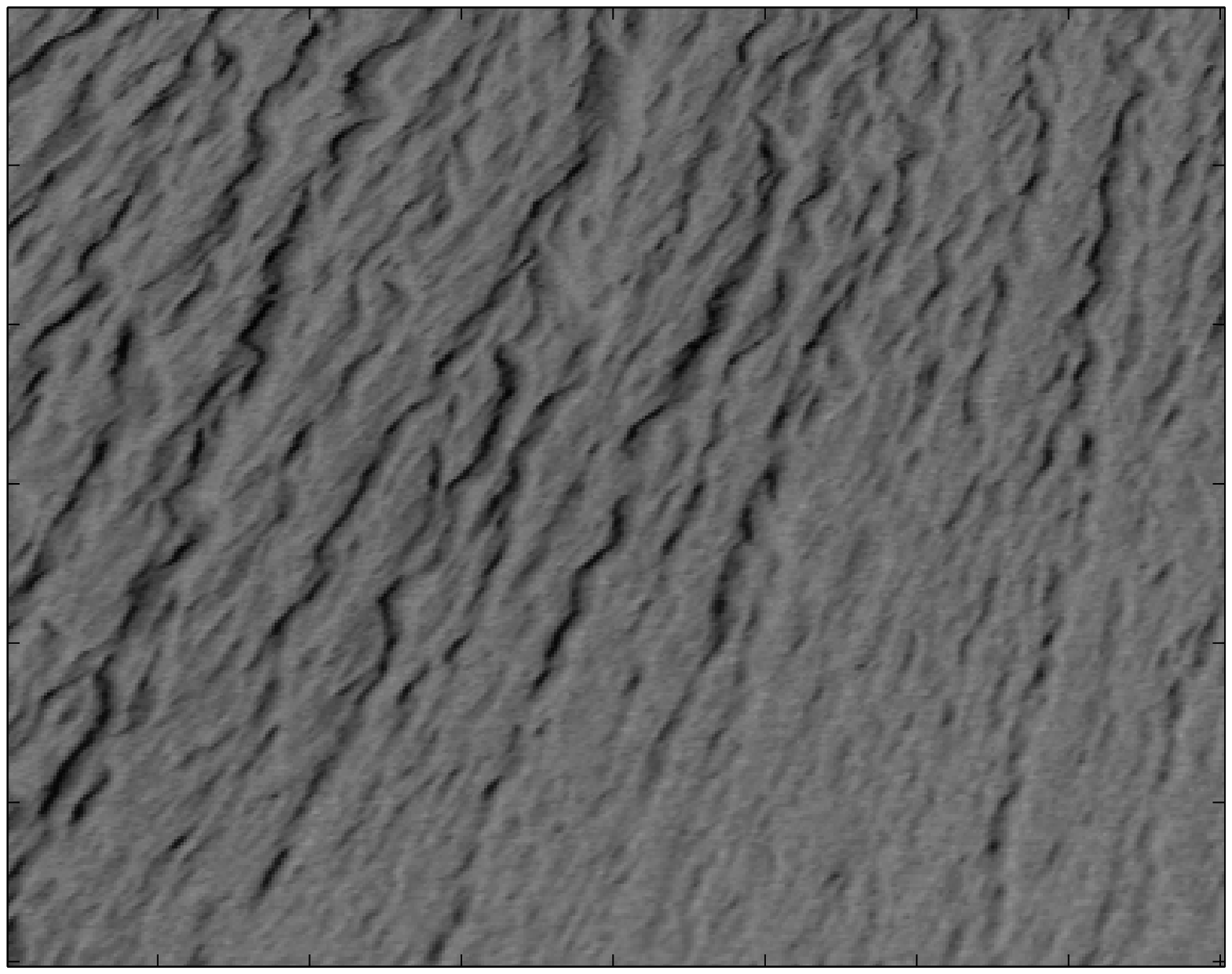}
\includegraphics[width=4cm]{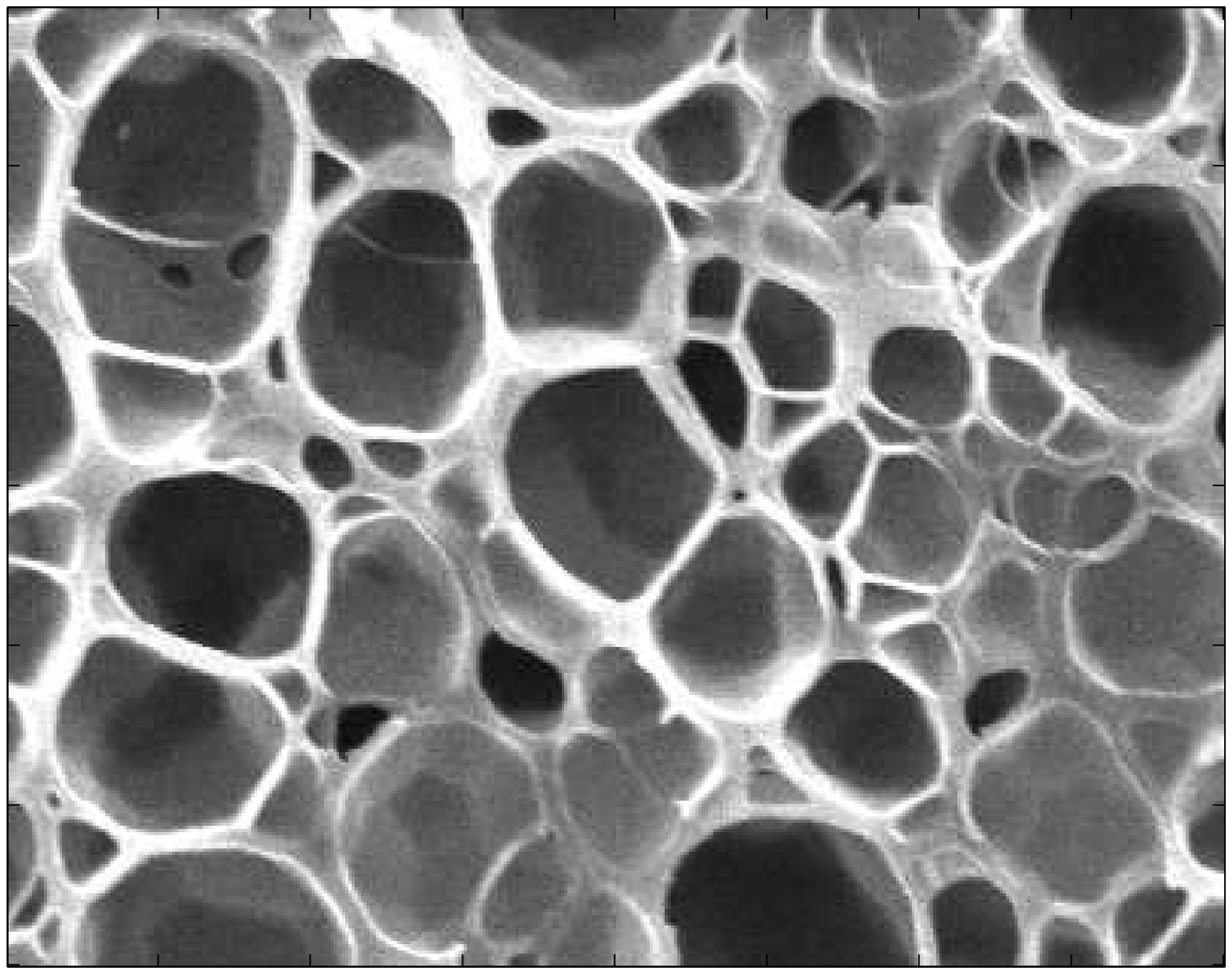}
\caption{Left: The image of the Yardangs region on the Mars. Right:
A scanning electron microscope picture of the surface of a
polyurethane sample foamed with supercritical carbon dioxide.}
 \label{Fig:Images}
\end{center}
\end{figure}

In this section we apply the generalized method to analyze two real
images, as shown in Fig.~\ref{Fig:Images}. Both pictures are
investigated by the MF-DFA approach since it contains automatically
the DFA analysis. The first example is the landscape image of the
Mars Yardangs region
\cite{AlvarezRamirez-Rodriguez-Cervantes-Carlos-2006-PA}, which can
be found at http://sse.jpl.nasa.gov. The size of the landscale image
is $2048\times1536$ pixels. The second example is a typical scanning
electron microscope (SEM) picture of the surface of a polyurethane
sample foamed with supercritical carbon dioxide. The size of the
foaming surface picture is $1200\times800$ pixels.

The SEM picture of the surface of a polyurethane sample were
prepared in an experiment of polymer foaming with supercritical
carbon dioxide. At the beginning of the experiment, several prepared
polyurethane samples were placed in a high-pressure vessel full of
supercritical carbon dioxide at saturation temperature for gas
sorption. After the samples were saturated with supercritical
${\rm{CO}}_2$, the carbon dioxide was quickly released from the
high-pressure vessel. Then the foamed polyurethane samples were put
into cool water to stabilize the structure cells. Pictures of the
foamed samples were taken by a scanning electron microscope.

The two images were stored in the computer as two-dimensional arrays
in 256 prey levels. We used Eq.~(\ref{Eq:eq30}) for the detrending
procedure. The two-dimensional arrays were investigated by the
multifractal detrended fluctuation analysis. For each picture, we
obtained the $\tau(q)$ function and the $h(q)$ function as well. If
$\tau(q)$ is nonlinear with respect to $q$ or, in other words,
$h(q)$ is dependent of $q$, then the investigated picture has the
nature of multifractality.

\subsection{Analyzing the Mars landscape image}

\begin{figure}[htb]
\begin{center}
\includegraphics[width=6.5cm]{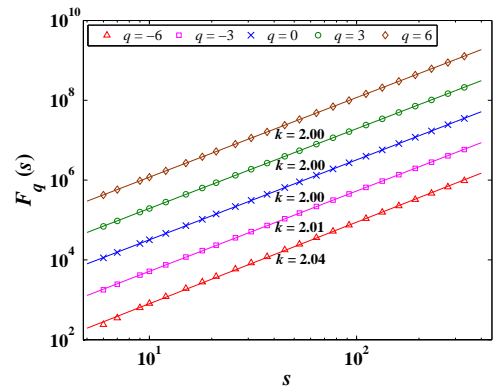}
\caption{Log-log plots of the detrended fluctuation function
$F_q(s)$ versus the lag scale $s$ for five different values of $q$.
The continuous lines are the best fits to the data. The plots for $q
= -3$, $q = 0$, $q = 3$, and $q = 6$ are shifted upwards for
clarity.} \label{Fig:Exmp:Mars:Fs}
\end{center}
\end{figure}

We first analyze the Mars landscape image shown in the left panel of
Fig.~\ref{Fig:Images} with MF-DFA. Figure \ref{Fig:Exmp:Mars:Fs}
illustrates the dependence of the detrended fluctuation $F_q(s)$ as
a function of the scale $s$ for different values of $q$ marked with
different symbols. The continuous curves are the best linear fits.
The perfect collapse of the data points on the linear lines
indicates the evident power law scaling between $F_q(s)$ and $s$,
which means that the Mars landscape is self-similar.

\begin{figure}[htb]
\begin{center}
\includegraphics[width=6.5cm]{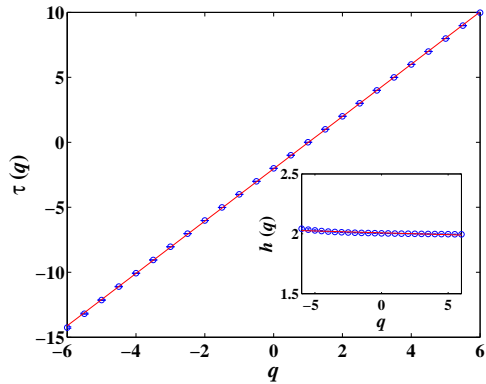}
\end{center}
\caption{Dependence of $\tau(q)$ with respect to $q$. The solid line
is the least squares fit to the data. The inset plots $h(q)$ as a
function of $q$.}
 \label{Fig:Exmp:Mars:tau}
\end{figure}

The slopes of the straight lines in Fig.~\ref{Fig:Exmp:Mars:Fs} give
the estimates of $h(q)$ and the function $\tau(q)$ can be calculated
accordingly. In Fig.~\ref{Fig:Exmp:Mars:tau} is shown the dependence
of $\tau(q)$ with respect to $q$ for $-6 \leqslant{q} \leqslant6$.
We observe that $\tau(q)$ is linear with respect to $q$. This
excellent linearity of $\tau(q)$ is consistent with the fact that
$h(q)$ is almost independent of $q$, as shown in the inset. Hence,
the Mars landscape image does not possess multifractal nature.

\subsection{Analyzing the foaming surface image}

\begin{figure}[htb]
\begin{center}
\includegraphics[width=6.5cm]{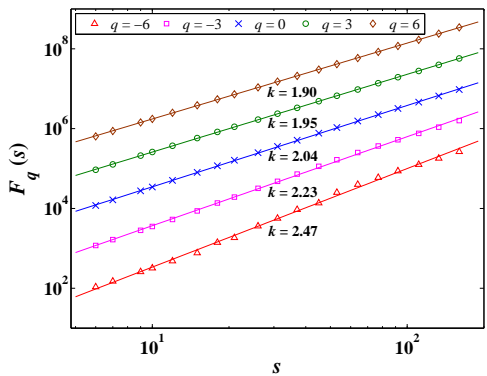}
\caption{Loglog plots of the detrended fluctuation function $F_q(s)$
versus the lag scale $s$ for five different values of $q$. The
continuous lines are the best fits to the data. The plots for $q =
-3$, $q = 0$, $q = 3$, and $q = 6$ are shifted upwards for clarity.}
\label{Fig:Exmp:Foam:Fs}
\end{center}
\end{figure}

Similarly, we analyzed the foaming surface shown in the right panel
of Fig.~\ref{Fig:Images} with the MF-DFA method. Figure
\ref{Fig:Exmp:Foam:Fs} illustrates the dependence of the detrended
fluctuation $F_q(s)$ as a function of the scale $s$ for different
values of $q$ marked with different symbols. The continuous curves
are the best linear fits. The perfect collapse of the data points on
the linear line indicates the evident power law scaling between
$F_q(s)$ and $s$, which means that the Foaming surface is
self-similar.

\begin{figure}[htb]
\begin{center}
\includegraphics[width=6.5cm]{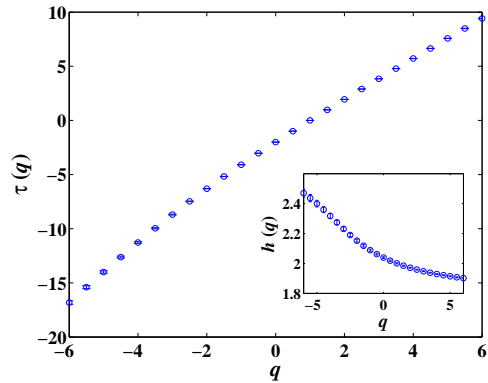}
\caption{Dependence of $\tau(q)$ with respect to $q$. The inset
shows $h(q)$ as a function of $q$.}
 \label{Fig:Exmp:Foam:tau}
\end{center}
\end{figure}

The values of $h(q)$ are estimated by the slopes of the straight
lines illustrated in Fig.~\ref{Fig:Exmp:Foam:Fs} for different
values of $q$. The corresponding function $\tau(q)$ is determined
according to Eq.~(\ref{Eq:MFDFA:tau}). In
Fig.~\ref{Fig:Exmp:Foam:tau} is illustrated $\tau(q)$ as a function
of $q$ for $-6 \leqslant{q} \leqslant6$. We observe that $\tau(q)$
is nonlinear with respect to $q$, which is further confirmed by the
fact that $h(q)$ is dependent of $q$, as shown in the inset. The
nonlinearity of $\tau(q)$ and $h(q)$ shows that the foaming surface
has multifractal nature.

\section{Discussion and conclusion}
\label{s1:Conclusion}

In summary, we have generalized the one-dimensional detrended
fluctuation analysis and multifractal detrended fluctuation analysis
to two-dimensional versions. Further generalization to higher
dimensions is straightforward. We have found that the
higher-dimensional DFA methods should be performed locally in the
sense that the cumulative summation should be conducted after the
partitioning of the higher-dimensional multifractal object.
Extensive numerical simulations validate our generalization. The
two-dimensional MF-DFA is applied to the analysis of the Mars
landscape image and foaming surface image. The Mars landscape is
found to be a fractal while the foaming surface exhibits
multifractality.

At last, we would like to stress that there are tremendous potential
applications of the generalized DFA in the analysis of fractals and
multifractals. In the two-dimensional case, the methods can be
adopted to the investigation of the roughness of fracture surfaces,
landscapes, clouds, and many other images possessing self-similar
properties. In the case of three dimensions, it could be utilized to
qualify the multifractal nature of temperature fields and
concentration fields. Possible examples in higher dimensions are
stranger attractors in nonlinear dynamics. Concrete applications
will be reported elsewhere in the future publications.

\begin{acknowledgements}
The SEM picture was kindly provided by Tao Liu. This work was
partially supported by National Basic Research Program of China
(Grant No. 2004CB217703), Fok Ying Tong Education Foundation (Grant
No. 101086), and Scientific Research Foundation for the Returned
Overseas Chinese Scholars, State Education Ministry of China.
\end{acknowledgements}

\bibliography{E:/papers/Bibliography}

\end{document}